\documentclass{article}

\begin{document}

\title{Scaling function and universal
amplitude combinations for self-avoiding polygons}

\author{C. Richard, A. J. Guttmann and I. Jensen\\
Department of Mathematics \& Statistics, \\
The University of Melbourne, Vic. 3010, Australia}

\maketitle

\begin{abstract}
%We derive the scaling function for self-avoiding polygons, 
%on the square and triangular lattices, enumerated by both perimeter 
%and area. It is the logarithm of an Airy function. 
We analyze new data for self-avoiding polygons, on the square and triangular lattices, 
enumerated by both perimeter and area, providing evidence that the scaling 
function is the logarithm of an Airy function.
The results imply universal amplitude combinations for all area moments and suggest
that rooted self-avoiding polygons may satisfy a $q$-algebraic functional equation.
\end{abstract}

\section{Introduction}

Self-avoiding polygons (SAP) on a lattice enumerated by perimeter are the
canonical model of ring polymers, while when enumerated by area as
well they are widely used as models of vesicles~\cite{LSF87,F89,FGW91}.
Despite their widespread applicability, remarkably little exact information
is known about their behaviour \cite{FGW91,CG93,C94,R00}, almost all available 
information being as a result of numerical studies~\cite{LSF87,FGW91,EG92,JG99,J00}. 
In this letter we give, for the first time, significant exact information,
subject to a very plausible conjecture. 
The validity of this conjecture is supported by abundant numerical evidence.

Consider the scaling behaviour of polygon models, whose two-variable generating 
functions are defined by 
\begin{equation}\label{pagf}
G(x,q) = \sum_{m,n} p_{m,n} x^m q^n,
\end{equation}
where $p_{m,n}$ is the number 
of polygons, equivalent up to a translation, with perimeter $m$, enclosing an area $n$,
on the triangular or square lattice. 
%Here $x$ is the variable conjugate to the perimeter, and $q$ is conjugate to the area.
Sometimes horizontal and vertical steps are distinguished, leading to a
3-variable generating function $G(x,y,q).$

These models, as well as simpler models, display 
scaling behaviour in the vicinity of the multicritical
point $((x,q) \rightarrow (x_c^-,1^-))$ of the form
\begin{equation}\label{sf1}
G(x,q) \sim G^{(reg)}(x,q) + (1 - q)^{\theta}F\left(\frac{x_c - x}{(1 - q)^\phi}\right),
\end{equation}
where 
$F(s) \sim {\rm const.}/s^\gamma$ as  $(s \rightarrow \infty)$ 
and $F(0) \sim {\rm const},$  
with $\gamma = -\frac{\theta}{\phi}.$ In this letter we derive and solve
the differential equation satisfied by the scaling function $F(s).$

In an earlier paper \cite{RG01} we discussed the scaling functions appropriate 
to very simple polygon models that satisfy $q$-linear functional equations,
of the form
\begin{equation}
G(x,q)=\sum_{k=1}^N a_k(x,q)G(q^kx,q) + b(x,q),
\end{equation}
where $a_k(x,q)$ and $b(x,q)$ are rational functions in $x$ and
$q$. 
In particular, for 
polygon models such as rectangles, Ferrers diagrams and stacks \cite{PO952}, 
the functional equation takes the simple form 
$$G_s(x,y,q) = \frac{y}{(1 - qx)^s} G_s(qx,y,q)+\frac{qxy}{1-qx},$$ 
where $s = 0, 1, 2$ 
characterizes rectangles, Ferrers diagrams and stacks respectively. 
This can be solved by iteration to give
\begin{equation}
G_s(x,y,q) = \sum_{n = 1}^\infty \frac{ q^n x y^n}{(1 - q^nx)(qx;q)^s_{n-1}},
\end{equation}
where $(t;q)_n = \prod_{k=0}^{n-1} (1 - q^kt)$ denotes the $q$-product.

For more complicated 
classes of polygons, such as staircase polygons, column-convex polygons, 
convex polygons, directed and convex polygons, the generating functions 
$G(x,q)$ are known (see \cite{BM95, BM96} and references therein). 
All these solvable cases, however, share the property that they are essentially 
one-dimensional, and so functional equations can be constructed that ``build up'' 
the polygon from smaller pieces. No such construction is possible for SAP.

For example, staircase polygons, which are defined by the rule 
that their perimeter consists of two directed paths, which proceed only in 
the positive $x$ and $y$ direction, share a common starting point and end point, 
and otherwise do not intersect. They satisfy the following non-linear functional 
equation \cite{PB95}:
\begin{eqnarray}\label{spfe}
G(x,y,q) = \frac{qxy}{1 - qx} + \frac{y + G(x,y,q)}{1 - qx} G(qx,y,q), 
\end{eqnarray}
which allows an iterative solution to be developed. 
It is
\begin{equation}\label{stair}
G(x,y,q) = qxy\frac{L(qx,qy)}{L(x,y)},
\end{equation}
where $L(x,y)$ is a $q$-deformed Bessel function
$$L(x,y) = \sum_{m,n \ge 0} \frac{(-1)^{n+m}x^n y^m q^{n+m+1 \choose
2}}{(q;q)_n(q;q)_m}.$$
The generating function by perimeter and area, $G(x,x,q)$ was first given in \cite{BG90}, while
the generating function by width and area, $G(x,1,q)$ was given in \cite{DF93}, 
in terms of $q$-Bessel functions. The full three-variable generating function was 
subsequently obtained \cite{BMV92}, and in the above $x-y$ symmetric form in \cite{BM95}.
In \cite{PB95} a semi-continuous version of
this model was solved, which has the same asymptotic behaviour
as the discrete model given here.
In \cite{P94}, the scaling function (\ref{sf1}) for the discrete model 
(with $\phi= \frac{2}{3}$ and $\theta = \frac{1}{3}$)
has been extracted from the solution (\ref{stair}).

Below we show how to derive the (non-linear) differential equation satisfied by the
scaling function for staircase polygons, using the method of dominant balance~\cite{RG01} 
on the quadratic functional equation~(\ref{spfe}).
It is
\begin{equation}~\label{F}
F(s)^2-4 f_1 F'(s)-f_0^2 s=0.
\end{equation}
%\begin{equation}\label{ss1}
%F(s)^2 + b F'(s) - c s = 0,
%\end{equation}
with $f_0^2=1/2$, $f_1=-1/32$ at the isotropic point $y_c=x_c=1/4.$ 
This is a special Riccati equation,
the solution being the logarithmic derivative of an Airy function.

Now, earlier numerical work \cite{PB95} prompted one of us \cite{PO95} to point out that
the SAP generating function for {\em rooted} self-avoiding polygons 
$G^{(r)}(x,q) = x\frac{d}{dx}G(x,q)$
has the same exponent values $\phi= \frac{2}{3}$ and $\theta = \frac{1}{3}$
as do staircase polygons, and to ask whether the scaling functions might
be the same. 
In this work we answer that question in the affirmative.
Of course, this is not to suggest that rooted SAP satisfy a similar, comparatively simple,
quadratic functional equation to that of staircase polygons. 

However let us tentatively 
assume that rooted SAP satisfy an $q$-algebraic functional
equation of {\em arbitrary} degree $N$,
\begin{equation}\label{qN}
\sum_{n=1}^N \sum_{k_1,\ldots,k_n} a_{k_1,\ldots,k_n}(x,q)
\prod_{i=1}^n G(q^{k_i}x,q) = b(x,q),
\end{equation}
where $a_{k_1,\ldots,k_n}(x,q)$ and $b(x,q)$ are polynomials in $x$ and $q$.
We require the sum to be finite.
Then, remarkably, by the method of dominant balance,
it follows that one is inexorably led to {\em the same}
differential equation for the scaling function~(\ref{F}) as found for staircase 
polygons, though with different constants $f_0$ and $f_1$. 
These two constants can be expressed in terms of the amplitude of the perimeter 
generating function $G(x,1)$ and the amplitude of the 
first area moment $A^{(1)}(x,1)=\sum_{m,n} n p_{m,n} x^m$
of the perimeter generating function.
Further, as we show below, the first area moment amplitude can be
given in terms of the critical point $x_c$, for any lattice!
From our extensive enumeration data~\cite{JG99} we have 
already estimated the required critical point and perimeter generating function 
amplitude accurately.

We can then solve this differential equation, and use it to predict the
amplitudes of all area weighted moments, 
$A^{(k)}(x,1)=\sum_{m,n}\frac{n!}{(n-k)!} p_{m,n} x^m,$
which may then be compared to our numerical evaluations, based on extensive
enumeration data, described below. We obtain complete agreement, up to the
accuracy of our numerical calculations for the first 10 area weighted moments,
which gives us sufficient confidence to conjecture that the scaling function
is exact.
Given the scaling function for rooted SAP, the scaling function for (unrooted) SAP is
obtained by integration.

\section{Scaling functions and $q$-algebraic polygon models}

The coefficients in the asymptotic expansion of the scaling function~(\ref{sf1})
are readily related to the area weighted moments of the perimeter and area 
generating function~(\ref{pagf}) by Taylor's theorem.
Expanding $G(x,q)$ about $q=1$ gives
\begin{equation}\label{expansion}
G(x,q) = \sum_{m,n} p_{m,n} x^m q^n = \sum_k g_k(x) (1-q)^k,
\end{equation}
where the functions $g_k(x)$ are proportional to the area weighted moments,
$g_k(x) = \frac{(-1)^k}{k!}A^{(k)}(x,1).$
If we make the common assumption that the singular behaviour of the coefficients is
\begin{equation}\label{expo}
g_k^{(sing)}(x) = \frac{f_k}{(x_c-x)^{\gamma_k}} + {\cal O}\left(
(x_c-x)^{-\gamma_k+1}\right),
\end{equation}
where $f_k$ is the amplitude,
then the assumption of the scaling form~(\ref{sf1}) constrains the exponents to be
$\gamma_k = \frac{k - \theta}{\phi}$ \cite{PB95},
while the amplitudes $f_k$ appear in the asymptotic expansion of the scaling
function \cite{RG01}
\begin{equation}\label{asym}
F(s) = \sum_{k=0}^\infty \frac{f_k}{s^{\gamma_k}}.
\end{equation}
The exponents $\theta$ and $\phi$ can be extracted from knowledge of the perimeter
generating function $g_0(x)$ and the first area weighted moment $g_1(x)$.
For staircase polygons, they can be computed recursively from the functional equation
(\ref{spfe}), giving $\theta=\frac{1}{3}$ and $\phi=\frac{2}{3}$ \cite{PB95}.

We next assume that the generating function $G(x,q)$ satisfies the 
$q$-algebraic functional equation~(\ref{qN}), and that the limit $q\to1$ leads 
to a non-trivial equation for the perimeter generating function $G(x,1)$ with a
singularity at $x=x_c$.
Assuming the existence of a scaling function (\ref{sf1}) about $x=x_c$ and $q=1$,
we show how its differential equation can be obtained from~(\ref{qN}).
For brevity, we state the argument under simplifying assumptions, which are sufficient
for the following numerical analysis.
The full argument will be presented elsewhere \cite{R01}.

Assume that the critical exponents $\theta=\frac{1}{3}$ and $\phi=\frac{2}{3}$ are 
known.
For rooted SAP, they can be obtained from numerical analysis of the perimeter
generating function and the first area weighted moment, as explained below. 
The method of dominant balance consists in expanding the functional equation about
$x=x_c$ and $q=1$, thereby introducing the scaling variable $s=(x_c-x)/(1-q)^\phi$.
In the scaling limit $q\to1$, this leads to a differential equation for the scaling
function $F(s)$.
If we assume the scaling form (\ref{sf1}), the generating function may be expanded
about the critical point as
\begin{equation}
G(q^kx,q) \sim G^{(reg)}((1-\epsilon)^k(x_c-s\epsilon^\phi),1-\epsilon) +
\epsilon^\theta
F(s) +
\epsilon^{1-\phi+\theta} k x_c F'(s) + {\cal O}
\left(\epsilon^{2-2\phi+\theta} \right),
\end{equation}
where $\epsilon=1-q$.
Observing the powers of $\epsilon$ that occur in each term, it follows that the 
$q$-algebraic functional equation (\ref{qN}) is a sum of terms the form
\begin{equation}
\left( \epsilon^\theta F \right)^{m_1} \left( \epsilon^{1-\phi+\theta} F' \right)^{m_2} 
\epsilon^{m_3} \left( s\epsilon^\phi\right)^{m_4}
\end{equation}
with non-negative integers $m_1,m_2,m_3,m_4$.
The differential equation results from taking the terms with {\it smallest} exponents in
$\epsilon$. 
Terms of order $\epsilon^0$ vanish, while terms of
${\cal O}\left(\epsilon^{1/3}\right)$ give a trivial equation and will therefore vanish, 
if a scaling function exists. 
Thus, the leading contribution in (\ref{qN}) in the scaling limit is of order 
$\epsilon^{2/3}$, which, by power counting results in the differential equation
(\ref{F}) for the scaling function.
%\begin{equation}~\label{F}
%F(s)^2-4 f_1 F'(s)-f_0^2 s=0.
%\end{equation}
An exactly solvable example is staircase polygons~(\ref{spfe}).

The solution of (\ref{F}) is uniquely determined by the asymptotic behaviour 
(\ref{asym}) and given by
\begin{equation}
F(s) = -4 f_1 \frac{d}{ds} \ln \mbox{Ai} \left( \left(\frac{f_0}{4
f_1}\right)^{2/3}s\right),
\end{equation}
where $\mbox{Ai}(x)=\frac{1}{\pi}\int_0^\infty\cos(t^3/3+tx) \, dt$ is the Airy function.
The coefficients $f_k$ in the asymptotic expansion~(\ref{asym}) of the scaling function
can be written 
\begin{equation}~\label{fk}
f_k = c_k f_1^k f_0^{1-k} \qquad (k=2,3,\ldots),
\end{equation}
where the numbers $c_k$ can be computed recursively from
\begin{equation}~\label{ck}
2 c_{k+1} + 6 k c_k + \sum_{l=2}^{k-1} c_l c_{k+1-l}=0 \qquad (k=2,3,\ldots).
\end{equation}
We have the initial values $c_1=1$ and $c_2=-5/2$.
The subsequent values are
$c_3= 15$, $c_4=-1105/8$, $c_5=1695$, $c_6=-414125/16$, $c_7=472200$, 
$c_8=-1282031525/128$, $c_9=242183775$, $c_{10}=-1683480621875/256$.

\section{Generation and analysis of series}

In order to test our predictions, we have generated the full perimeter
and area generating function for square-lattice SAPs up to perimeter 66
and for triangular-lattice SAPs up to perimeter 29. For square-lattice
SAPs, we have generated the first ten area weighted moments of the 
perimeter generating function up to perimeter 86.
The series are available upon request.
The method used to enumerate square-lattice SAP is an
enhancement of the method devised by Enting \cite{Ent} in his
pioneering work. 
Recent improvements to the method are developed and described in~\cite{JG99,J00}.
For the triangular lattice, the method and program used is precisely that described 
in \cite{EG92}.

In \cite{CG93} it was argued that self-avoiding polygons should admit certain universal
combinations between the amplitudes of the area weighted moments.
Here, we give them all in terms of the coefficient amplitudes $A_k$ of the functions 
$g_k(x)$ in the expansion (\ref{expansion}).
The coefficients of the functions $g_k(x)$ grow asymptotically as
\begin{equation}
[x^n]g_k(x) \simeq \sigma A_k x_c^{-n} n^{\gamma_k-1},
\end{equation} 
where $\sigma$ is a constant such that $p_{m,n}$ is nonzero only if $m$ is divisible 
by $\sigma$.
Thus $\sigma=2$ for the square lattice and $\sigma=1$ for the triangular lattice 
\cite{CG93}.
The amplitudes $A_k$ are related to the amplitudes $f_k$ in (\ref{expo}) via
\begin{equation}~\label{Ak}
A_k = \frac{f_k}{x_c^{\gamma_k} \Gamma(\gamma_k)}.
\end{equation}
By definition, the absolute values of these amplitudes coincide for rooted SAP and for 
unrooted SAP.
The conjecture in \cite{CG93} reads in our notation that all products 
$A_k A_0^{k-1}\sigma^k$ are universal.
In particular, from amplitude relations given in \cite{CG93} and in \cite{C94} we
infer $A_1\sigma=-\frac{1}{4\pi}$, independent of the lattice!
Using this and the assumed scaling equation, we obtain from~(\ref{fk}) and~(\ref{Ak}) 
for even $k=2m$
\begin{eqnarray}
A_{2m} A_0^{2m-1} &=& -c_{2m} A_1^{2m} \frac{(3m-2)!}{(6m-3)!}\frac{2^{4m-2}}{\pi^m}
\nonumber \\
&=& -\frac{c_{2m}}{4\pi^{3m}\sigma^{2m}}\frac{(3m-2)!}{(6m-3)!}, 
\end{eqnarray}
where the $c_k$ are given by~(\ref{ck}).
For odd $k=2m+1$, we obtain
\begin{eqnarray}
A_{2m+1} A_0^{2m} &=& c_{2m+1} A_1^{2m+1} \frac{1}{(3m)! 2^{2m}\pi^m} \nonumber\\
&=& -\frac{c_{2m+1}}{(3m)!\pi^{3m+1}2^{6m+2}\sigma^{2m+1}}.
\end{eqnarray}

We now test these predictions against numerical data obtained for rooted SAP on
the square and triangular lattices. Note that this tests our entire theory, as
these amplitude relations follow from the scaling function we derived.
A numerical analysis of the singularity of the area weighted moments
confirms the behaviour assumed in~(\ref{expo}) with $\theta=\frac{1}{3}$
and $\phi=\frac{2}{3}$ within 
accuracy given by series data.
As argued above, we obtain estimates of the coefficients of the scaling function 
from estimates of the amplitudes of the area weighted moments.
These amplitudes are estimated by a direct fit to the expected asymptotic form.
That is to say, we fit the the coefficients to the assumed form
$ [x^n]g_k(x) \approx x_c^{-n} n^{\gamma_k-1}\sum_{i\ge 0}a_i/n^{f(i)}.$
If there is no non-analytic correction term, then $f(i)=i.$
A square-root correction term means $f(i)=i/2.$
In all cases, our procedure is to {\em assume} a particular
form for $f(i),$ and observe the extent to which it fits the data. With the long
series we now have at our disposal, it is usually easy to see
if the wrong assumption has been made---the sequence of
amplitude estimates $a_i$ either diverges to infinity
or converges to zero.
Once the correct assumption is made, convergence is usually
rapid and obvious. Detailed demonstrations of the method can
be found in \cite{CG96,JG99}. In this way, we obtain the following
results for the leading amplitudes:
\newline

\begin{tabular}{llll}
\hline \hline
amplitude & exact value & square lattice & triangular lattice \\
\hline
$-A_1 \sigma$ &.79578$\times 10^{-1}$  
&.7957(1)$\times 10^{-1}$  
&.796(1)$\times 10^{-1}$ \\
$A_2 A_0 \sigma^2$ & .33595$\times 10^{-2}$ 
&.3359(1)$\times 10^{-2}$ 
& .335(1)$\times 10^{-2}$\\
$-A_3 A_0^2 \sigma^3$ 
& .10025$\times 10^{-3}$
&.1002(1)$\times 10^{-3}$ 
& .100(1)$\times 10^{-3}$ \\
$A_4 A_0^3 \sigma^4$ 
& .23755$\times 10^{-5}$
& .2375(1)$\times 10^{-5}$
& .237(1)$\times 10^{-5}$ \\
$-A_5 A_0^4\sigma^5$ 
&.47574$\times 10^{-7}$
&.4757(1)$\times 10^{-7}$
&.475(1)$\times 10^{-7}$ \\
$A_6 A_0^5\sigma^6$ 
&.83663$\times 10^{-9}$
&.8366(1)$\times 10^{-9}$
& .83(1)$\times 10^{-9}$ \\
$-A_7 A_0^6\sigma^7$ 
&.13251$\times 10^{-10}$
&.1325(1)$\times 10^{-10}$ 
& .13(1)$\times 10^{-10}$ \\
$A_8 A_0^7\sigma^8$ 
& .19242$\times 10^{-12}$ 
&.1924(1)$\times 10^{-12}$
& .18(1)$\times 10^{-12}$ \\
$-A_9 A_0^8\sigma^9$ 
&.25947$\times 10^{-14}$ 
&.2594(1)$\times 10^{-14}$
&.25(1)$\times 10^{-14}$ \\
$A_{10} A_0^9\sigma^{10}$ 
&.32806$\times 10^{-16}$
&.3280(1)$\times 10^{-16}$
& .31(1)$\times 10^{-16}$\\
\hline \hline
\end{tabular} 
\newline
\newline
It is clear that the results for the first 10 area weighted moments
are in excellent agreement with the numerical estimates. On
this basis we conjecture that the scaling function is correct.
Accepting this conjecture, it follows that
the scaling function $F^{(r)}(s)$ for rooted SAP is
\begin{equation}\label{scalfct}
F^{(r)}(s) = - \frac{x_c}{\pi\sigma} \frac{d}{ds} \ln \mbox{Ai} \left(
 \frac{\pi}{x_c}  \left(2\sigma A_0\right)^{2/3}    s\right),
\end{equation}
where we have $x_c=0.379052277(1)$, $\sigma=2$, $A_0=0.2811506(1)$ for the square 
lattice and $x_c=0.240917(1)$, $\sigma=1$, $A_0=0.263936(1)$ for the triangular 
lattice.
We can check this result further by comparing $F^{(r)}(0)$ with series data evaluated at
$x=x_c$, according to (\ref{sf1}).
We predict from (\ref{scalfct}) the values $F^{(r)}(0)=0.394188(1)$ for the square 
lattice and $F^{(r)}(0)=0.476161(1)$ for the triangular lattice, which is in agreement 
with the numerical estimates $F^{(r)}(0)=0.39(1)$ (square lattice) and 
$F^{(r)}(0)=0.47(1)$ (triangular lattice). 
The scaling function $F(s)$ of (unrooted) SAP follows from (\ref{scalfct}) as
\begin{equation}
F(s) =  \frac{1}{\pi\sigma} \ln \mbox{Ai} \left(
   \frac{\pi}{x_c}  \left(2\sigma A_0\right)^{2/3}   s\right)
\end{equation}
with exponents $\theta=1$ and $\phi=\frac{2}{3}$. 
\section{ Conclusion}
We have given what we believe to be the exact scaling functions
for self-avoiding polygons in the vicinity of the tricritical point,
for both the square and triangular lattices.
The result follows from the assumption that the two-variable generating
function for rooted self-avoiding polygons satisfies a $q$-algebraic functional 
equation. 
It is possible that even more striking results could flow from exploration of this
assumption, a topic we are currently investigating.
A corresponding result for the scaling function
in the inflated regime, $q > 1$ is given in \cite{PO99},
though the SAP structure is largely washed out in this regime.

\section{Acknowledgements}
We would like to thank Ian Enting for resurrecting his
old triangular lattice program which we used to produce triangular lattice data.
We thank Thomas Prellberg for clarifying discussions.
Two of us, AJG and IJ would like to thank the Australian Research Council
for continued support, while CR would like to thank the German Science Foundation
(DFG) similarly.

\end{document}